\documentclass[article]{JFM-FLM_Au_mod}
\usepackage[dvipsnames]{xcolor}
\usepackage{mathtools}
\usepackage{svg}
\usepackage{subcaption}

\usepackage{graphicx}
\usepackage{hyperref}
\hypersetup{colorlinks=true,citecolor=blue}

\lefttitle{R. van Buel, B. Cichocki and J. C. Everts}
\righttitle{Journal of Fluid Mechanics}

\title{Linear odd electrophoresis of a sphere in a charged chiral active fluid}

\author{Reinier van Buel\aff{1}, Bogdan Cichocki\aff{1} \and Jeffrey C. Everts\aff{1,2}}

\affiliation{\aff{1}Institute of Theoretical Physics, Faculty of Physics, University of Warsaw, Pasteura 5, 02-093 Warsaw, Poland
\aff{2}Institute of Physical Chemistry, Polish Academy of Sciences, 01-224 Warsaw, Poland}

\corresau{Jeffrey C. Everts, \email{jeffrey.everts@fuw.edu.pl}}

\def\*#1{\mathbf{#1}}
\def\!#1{\mathbf{\hat#1}}

\begin{document}

\maketitle
\thispagestyle{empty}

\begin{abstract}
The electrophoresis of charged colloidal particles in fluids exhibiting odd viscosity represents a fundamental challenge in understanding transport phenomena within charge-stabilised chiral active suspensions. Here, we consider a charged chiral active fluid, where electrokinetics is coupled to odd Stokes flow, to explore how classical results from electrophoresis in Newtonian fluids are modified in the presence of odd viscosity.
In particular, we derive a general expression for the electrophoretic mobility for particles of any shape, under weak external electric fields, using the Lorentz reciprocal theorem for odd fluids. By applying this result to a charged sphere at low zeta potentials, we obtain an exact, closed-form analytical expression for the electrophoretic mobility, valid for arbitrary values of the Debye screening length and the odd-viscosity coefficient. 
Similar to Newtonian fluids, we find that the electrophoretic mobility is proportional to the translational mobility of an uncharged sphere, modulated by the Henry function. However, unlike in Newtonian fluids, odd viscosity leads to directional asymmetries in the electrophoretic mobility tensor that persist even for thin electric double layers. This case contrasts significantly with a charged anisotropic particle suspended in an isotropic Newtonian fluid, where anisotropic effects would vanish under the same electrostatic-screening conditions.
\end{abstract}

\begin{keywords}
\end{keywords}


\section{Introduction}
\label{sec:introduction}
Chiral active fluids {are characterised by a non-vanishing} spin-angular momentum density that results in odd contributions to their transport coefficients \citep{julicher2018}. 
One example that has attracted significant interest in recent years 
stems from antisymmetric contributions to the viscosity tensor,
called odd viscosity \citep{Avron:1988, Banerjee:2017, Fruchart:2023}.
The effects of odd viscosity on fluid flow have been well documented in recent theoretical work: analytically exact solutions have been obtained for the hydrodynamic flow profiles of two-dimensional rigid disks in compressible fluids with odd viscosity \citep{Hosaka:2025}, of a spherical particle in a three-dimensional unbounded flow \citep{Hosaka:2024,Everts:2024, Everts:2024b}, and of bubbles 
\citep{Khain:2022}.
In all cases, the odd viscosity leads to azimuthal components in the flow field, even at low (but non-vanishing) Reynolds numbers \citep{Lier:2024} or in compressible fluids \citep{Lier:2023}. These results have also been generalised to many particles suspended in fluids with small odd viscosity \citep{yuan2023}.

Despite the significant theoretical progress on odd hydrodynamics, the majority of experiments are restricted to two spatial dimensions, such as spinning colloids under the effect of a rotating magnetic field \citep{Soni:2019} and electrons in graphene subjected to a magnetic field \citep{berdyugin2019}. In three dimensions, the experimental realisations of odd viscosity are sparse, with the best-known example being
magnetised polyatomic gases \citep{beenakker1970}. Here, the effects of odd viscosity are, however, much smaller than those caused by the shear viscosity. 
A promising candidate to observe significant three-dimensional odd viscosity is the recent work of \citet{chen2025}, where suspended cylindrical particles are magnetically rotated at intermediate Reynolds numbers. However, there is still a major challenge in obtaining fluid-like behaviour of such suspensions.
Here, we expect charge stabilisation to be vital in enabling such experiments, to mitigate the effects of irreversible aggregation of the particles due to attractive van der Waals forces \citep{verwey1955}. 

Anticipating on the importance of charge stabilisation for such cases, we introduce the notion of a charged chiral active fluid. As a concrete example, a charge-stabilised ferrofluid \citep{Oehlsen:2022} driven by a rotating magnetic field could serve as a potential experimental realisation. Further promising candidates are motor proteins, as they are rotating structures that are naturally charged because of their amino-acid sequence \citep{Ito:2009}. A key fundamental phenomenon in such charged systems is electrophoresis --the motion of suspended charged particles under the influence of an external electric field \citep{Ohshima:2006}, which will be the focus in this work.

Here, our aim is to generalise certain classical results for electrophoresis in (isotropic) Newtonian fluids to odd fluids pertaining to the case of weak external electric fields, known as linear electrophoresis. 
For isotropic systems and thin electric double layers, 
the electrophoretic mobility no longer depends on the shape or size of the suspended particle, as shown by \citet{Smoluchowski} [for an English translation, see \citet{Cichockibook}].
The opposite limit of large screening lengths is called the Hückel limit, for which the first analytical solution was derived for a uniformly charged sphere \citep{Huckel}.
For intermediate Debye screening lengths and small zeta 
potentials, the electrophoresis of a sphere is governed by the Henry equation \citep{Henry:1931}, 
which interpolates the Hückel and Smoluchowski limits.
Odd viscosity introduces fluid anisotropy and antisymmetric contributions to the hydrodynamic drag, 
yet,
their influence on electrophoresis is not known, and this is an important outstanding problem.

In this paper, we characterise the linear electrophoresis of a charged colloidal particle in fluids with odd viscosity.
We derive a general expression for the electrophoretic mobility for particles of any shape by applying the Lorentz reciprocal theorem for odd fluids \citep{Vilfan:2023}.
We apply this expression to analyse the electrophoretic mobility tensors in the presence of odd viscosity within the Henry approximation, and in the Hückel and Smoluchowski limits.  Within this regime, we find an exact analytical result for the electrophoretic mobility of a charged spherical particle with uniform zeta potential for arbitrary Debye screening lengths, by using the known exact flow field of an uncharged sphere in an odd fluid \citep{Meissner:2025}.

\begin{figure}
    \centering
    \includegraphics[width=0.45\linewidth]{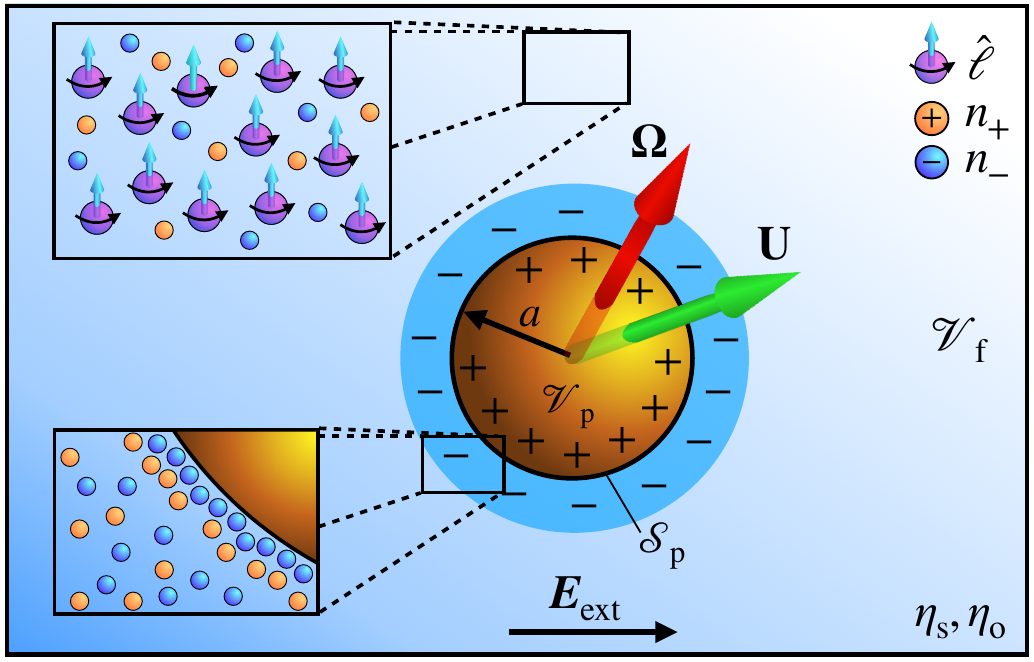}
    \caption{
    Schematic of a charged spherical particle with radius $a$, translating with velocity $\boldsymbol{U}$, and rotating with angular velocity $\boldsymbol{\Omega}$ in an electrolyte solution with shear viscosity $\eta_\mathrm{s}$, odd viscosity $\eta_\mathrm{o}$, cation number density $n_+$ (orange) and anion number density $n_-$ (blue) under the influence of an external field $\boldsymbol{E}_\mathrm{ext}$. The volume of the particle is $\mathcal{V}_\mathrm{p}$, and the volume of the fluid is $\mathcal{V}_\mathrm{f}$. The top inset depicts self-spinning (active) particles with intrinsic angular momentum $\boldsymbol{\hat{\ell}}$ (purple) that leads to the odd viscosity contribution. The bottom inset shows the electric double layer formed around a positively charged particle. For simplicity, a binary monovalent electrolyte has been depicted.
    }
    \label{fig:schematic}
\end{figure}

\section{Odd Poisson-Nernst-Planck-Stokes equations}
We consider an unbounded incompressible odd fluid, which is steady and quiescent. The fluid contains $N$ species of ions with number densities $n_i(\boldsymbol{r})$, for $i=1,...,N$, and total charge density $\rho(\boldsymbol{r})=\sum_{i=1}^Nz_i e n_i(\boldsymbol{r})$. Here, $z_i$ is the valency of ion species $i$, and $e$ is the elementary charge unit. In this charged chiral active fluid with domain $\mathcal{V}_\mathrm{f}$, we consider a rigid charged particle occupying 
a volume $\mathcal{V}_\mathrm{p}$, which is impenetrable 
to
the free ions in the solution. Neglecting magnetic effects, the local electric field can be expressed as $\boldsymbol{E}(\boldsymbol{r})=-\nabla\psi(\boldsymbol{r})$,
and the electrostatic potential $\psi(\boldsymbol{r})$ satisfies 
\begin{equation}
\varepsilon\nabla^2\psi(\boldsymbol{r})=-\rho(\boldsymbol{r}), \quad \boldsymbol{r}\in\mathcal{V}_\mathrm{f}, \quad \nabla^2\psi(\boldsymbol{r})=0, \quad \boldsymbol{r}\in\mathcal{V}_\mathrm{p}, \label{eq:Poisson}
\end{equation}
where $\varepsilon$ is the homogeneous dielectric constant of the solvent. 
In general,
the dielectric properties 
of an odd fluid
are described by the dielectric tensor $\boldsymbol{\varepsilon}(\boldsymbol{\hat\ell})$ with tensorial structure $\boldsymbol{\varepsilon}(\boldsymbol{\hat\ell})=\varepsilon_\parallel\boldsymbol{\hat\ell}\boldsymbol{\hat\ell}+\varepsilon_\perp(\mathsfbi{I}-\boldsymbol{\hat\ell}\boldsymbol{\hat\ell})+\varepsilon_\mathrm{o}(\boldsymbol{\epsilon}\cdot\boldsymbol{\hat\ell})$. Here, we assume that the self-spinning particles, yielding $\boldsymbol{\ell}\neq\boldsymbol{0}$,
are not polarisable,
and their interaction with the solvent is isotropic, 
and therefore to a good approximation, $\varepsilon=\varepsilon_\parallel=\varepsilon_\perp$ and $\varepsilon_\mathrm{o}=0$. 

A schematic of the particle and the fluid is given in Fig.~\ref{fig:schematic}.
The whole system is subjected to an external electric field $\boldsymbol{E}_\mathrm{ext}$, which results in motion of the suspended particle.  To compute this motion, we need to find the relation between $\rho(\boldsymbol{r})$ and the fluid-velocity field $\boldsymbol{v}(\boldsymbol{r})$. 
Therefore, we consider the total stress tensor $\boldsymbol{\sigma}(\boldsymbol{r};\boldsymbol{\hat\ell})=\boldsymbol{\sigma}_\mathrm{H}(\boldsymbol{r};\boldsymbol{\hat\ell})+\boldsymbol{\sigma}_\mathrm{E}(\boldsymbol{r})$,  
which consists of a hydrodynamic part
\begin{gather}
\boldsymbol{\sigma}_\mathrm{H}(\boldsymbol{r};\boldsymbol{\hat\ell})=-p(\boldsymbol{r})\mathsfbi{I}+2\eta_\mathrm{s}\mathsfbi{e}(\boldsymbol{r})+2\eta_\mathrm{o}[\mathsfbi{e}(\boldsymbol{r})\cdot\boldsymbol{\epsilon}\cdot\boldsymbol{\hat{\ell}}+\boldsymbol{\hat{\ell}}\cdot\boldsymbol{\epsilon}\cdot\mathsfbi{e}(\boldsymbol{r})], \quad \boldsymbol{r}\in\mathcal{V}_\mathrm{f} \, , \label{eq:hyd} 
\end{gather}
and the electrostatic Maxwell stress tensor
\begin{gather}
\boldsymbol{\sigma}_\mathrm{E}(\boldsymbol{r})=\varepsilon\left[\boldsymbol{E}(\boldsymbol{r})\boldsymbol{E}(\boldsymbol{r})-\frac{1}{2}{E}(\boldsymbol{r})^2\mathsfbi{I}\right],\quad \boldsymbol{r}\in\mathcal{V}_\mathrm{f} \, . \label{eq:Maxwell}
\end{gather}
Here, $p(\boldsymbol{r})$ is the absolute pressure, and $e_{\alpha\beta}(\boldsymbol{r})=[\partial_\alpha v_\beta(\boldsymbol{r})+\partial_\beta v_\alpha(\boldsymbol{r})]/2$ is the strain-rate tensor. The fluid is characterised by a dynamic shear viscosity $\eta_\mathrm{s}$ and an intrinsic spin angular momentum $\boldsymbol{\ell}=4\eta_\mathrm{o}\boldsymbol{\hat{\ell}}$, which in a spatially uniform non-equilibrium steady state has a magnitude proportional to the odd-viscosity coefficient $\eta_\mathrm{o}$ \citep{Banerjee:2017,Markovich:2021}.
In the creeping-flow regime, the balance of linear momentum is given by $\partial_\beta\sigma_{\alpha\beta}(\boldsymbol{r};\boldsymbol{\hat\ell})=0$, with Greek indices running over Cartesian coordinates $\{x,y,z\}$ and an implied summation when the indices are repeated. Together with the incompressibility condition and Eq. \eqref{eq:Poisson}, we find for $\boldsymbol{r}\in\mathcal{V}_\mathrm{f}$ that
\begin{gather}
\eta_\mathrm{s}\nabla^2\boldsymbol{v}(\boldsymbol{r})-\nabla\tilde{p}(\boldsymbol{r})+\eta_\mathrm{o}(\boldsymbol{\hat\ell}\cdot\nabla)[\nabla\times\boldsymbol{v}(\boldsymbol{r})]=-\rho(\boldsymbol{r})\boldsymbol{E}(\boldsymbol{r}), \quad
\nabla\cdot\boldsymbol{v}(\boldsymbol{r})=0, \label{eq:linmom}
\end{gather}
with effective pressure $\tilde{p}(\boldsymbol{r})=p(\boldsymbol{r})+2\eta_\mathrm{o}\boldsymbol{\hat\ell}\cdot[\nabla\times\boldsymbol{v}(\boldsymbol{r})]$. 
Next, the steady flux of free ions $\boldsymbol{J}_i(\boldsymbol{r})$ is conserved, i.e. $\nabla\cdot\boldsymbol{J}_i(\boldsymbol{r})=0$, where 
\begin{gather}
\boldsymbol{J}_i(\boldsymbol{r})=-\mathsfbi{D}_i(\boldsymbol{\hat\ell})\cdot\left[\nabla n_i(\boldsymbol{r})+\frac{z_ie}{k_\mathrm{B}T}n_i(\boldsymbol{r})\nabla\psi(\boldsymbol{r})\right]+n_i(\boldsymbol{r})\boldsymbol{v}(\boldsymbol{r}), \quad \boldsymbol{r}\in\mathcal{V}_\mathrm{f}, \quad i=1,...,N,\label{eq:ionflux}
\end{gather}
with $T$ the temperature, and $k_\mathrm{B}$ the Boltzmann constant. The terms in Eq. \eqref{eq:ionflux} describe ionic diffusion, electromigration, and advection, respectively. Furthermore, $\mathsfbi{D}_i(\boldsymbol{\hat\ell})$ are the {ionic} diffusion tensors with a tensorial structure similar to the dielectric tensor. However, in this work its precise form is irrelevant, as we shall see later in Sec.~\ref{sec3}. Eqs. \eqref{eq:Poisson}, \eqref{eq:linmom}, and \eqref{eq:ionflux} constitute the simplest generalisation of the Poisson-Nernst-Planck-Stokes equations \citep{Hunter} to odd fluids. To the best of our knowledge, we are the first to introduce such a notion of odd electrokinetics.
 
Our goal is to analyse the steady motion of an ion-impenetrable particle
with a no-slip surface $\mathcal{S}_\mathrm{p}$, described by
\begin{equation}
\boldsymbol{v}(\boldsymbol{r})=\boldsymbol{U}+\boldsymbol{\Omega}\times\boldsymbol{r}, \quad \boldsymbol{\hat n}(\boldsymbol{r})\cdot\boldsymbol{J}_i(\boldsymbol{r})=0, \quad i=1,...,N, \quad \boldsymbol{r}\in\mathcal{S}_\mathrm{p}, \label{eq:bc1}
\end{equation}
where $\boldsymbol{U}$ is 
its
 translational velocity, and $\boldsymbol{\Omega}$ is
 its
rotational velocity. Henceforth, $\boldsymbol{\hat n}(\boldsymbol{r})$ is a normal 
vector
pointing from the fluid to particle, with a caret denoting normalised vectors. We note that assessing the location of the slip plane can be quite complex in a charged odd fluid due to the presence of solvent, ionic, and spinner degrees of freedom. Here, we view the location and existence of the slip plane as a given (experimental) input. Electrostatic boundary conditions depend on the charge functionality of the particle (insulating, charge-regulating, or conducting) and will be specified later. Far-field conditions are
\begin{equation}
\lim_{r\rightarrow\infty}\boldsymbol{v}(\boldsymbol{r})=\boldsymbol{0},\quad \lim_{r\rightarrow\infty}\psi(\boldsymbol{r})/r=-\boldsymbol{E}_\mathrm{ext}\cdot\boldsymbol{\hat r}, \quad \lim_{r\rightarrow\infty}n_i(\boldsymbol{r})=n_i^\infty, \quad i=1,...,N, \label{eq:bc2}
\end{equation}
where the $n_i^\infty$ are constant bulk ion densities satisfying local charge neutrality $\sum_i z_in_i^\infty=0$. The forces $\boldsymbol{F}_\mathrm{H,E}$ and torques $\boldsymbol{T}_\mathrm{H,E}$ that the particle exerts on the charged chiral active fluid are
\begin{equation}
\boldsymbol{F}_j=\int_{\mathcal{S}_\mathrm{p}}\, \mathrm{d}S\, \boldsymbol{\sigma}_j(\boldsymbol{r})\cdot\boldsymbol{\hat n}(\boldsymbol{r}), \quad \boldsymbol{T}_j=\int_{\mathcal{S}_\mathrm{p}}\, \mathrm{d}S\, \boldsymbol{r}\times[\boldsymbol{\sigma}_j(\boldsymbol{r})\cdot\boldsymbol{\hat n}(\boldsymbol{r})], \quad j=\mathrm{H,E}.
\end{equation}
Since the particle is subjected to steady motion, we have $\boldsymbol{F}_\mathrm{H}+\boldsymbol{F}_\mathrm{E}=\boldsymbol{0}$ and $\boldsymbol{T}_\mathrm{H}+\boldsymbol{T}_\mathrm{E}=\boldsymbol{0}$.
We are interested in finding the relation between the translational velocity $\boldsymbol{U}$ and rotational velocity $\boldsymbol{\Omega}$ with $\boldsymbol{E}_\mathrm{ext}$, given by 
\begin{equation}
\boldsymbol{U}=\boldsymbol{\mu}^\mathrm{tE}(\boldsymbol{\hat\ell})\cdot\boldsymbol{E}_\mathrm{ext}, \quad \boldsymbol{\Omega}=\boldsymbol{\mu}^\mathrm{rE}(\boldsymbol{\hat\ell})\cdot\boldsymbol{E}_\mathrm{ext}. \label{eq:eletensors}
\end{equation}
These relations define the electrophoretic mobility tensor $\boldsymbol{\mu}^\mathrm{tE}(\boldsymbol{\hat\ell})$ and the electrorotation tensor $\boldsymbol{\mu}^\mathrm{rE}(\boldsymbol{\hat\ell})$, which typically depend on the direction of intrinsic spin-momentum of the fluid ${\boldsymbol{\hat\ell}}$. Here, we want to compute these tensors for small external electric fields -- known in the literature as linear electrophoresis -- where both tensors do not depend on $\boldsymbol{E}_\mathrm{ext}$.

\section{General result for electrophoretic mobility}
\label{sec3}
For small electric fields, the equations are effectively linear, and we can use the Lorentz reciprocal theorem to find expressions for $\boldsymbol{\mu}^\mathrm{tE}(\boldsymbol{\hat\ell})$ and $\boldsymbol{\mu}^\mathrm{rE}(\boldsymbol{\hat\ell})$ by generalising the considerations of \citet{Teubner:1982} to the odd case. We consider an incompressible flow of interest $\boldsymbol{v}(\boldsymbol{r})$ and an incompressible auxiliary flow $\boldsymbol{v}^{(0)}(\boldsymbol{r})$ satisfying the same constitutive relation Eq. \eqref{eq:hyd}, but with different body forces $\boldsymbol{f}(\boldsymbol{r})$ and $\boldsymbol{f}^{(0)}(\boldsymbol{r})$, respectively, acting on the fluid:
\begin{equation}
\partial_\beta\sigma_{\mathrm{H},\alpha\beta}(\boldsymbol{r};\boldsymbol{\hat\ell})+f_\alpha(\boldsymbol{r})=0, \quad \partial_\beta\sigma_{\mathrm{H},\alpha\beta}^{(0)}(\boldsymbol{r};\boldsymbol{\hat\ell})+f_\alpha^{(0)}(\boldsymbol{r})=0.
\end{equation}
Both flows do not necessarily satisfy the same boundary conditions and are related
by the ``odd'' Lorentz reciprocal theorem \citep{Vilfan:2023}, 
resulting 
from microscopic time reversibility, which leads to
\begin{gather}
\oint_{\mathcal{S}_\mathrm{p}}\mathrm{d}S\, \boldsymbol{v}^{(0)}(\boldsymbol{r};-\boldsymbol{\hat\ell})\cdot\boldsymbol{\sigma}_\mathrm{H}(\boldsymbol{r};\boldsymbol{\hat\ell})\cdot\boldsymbol{\hat n}(\boldsymbol{r})+\int_{\mathcal{V}_\mathrm{f}}\mathrm{d}V\, \boldsymbol{v}^{(0)}(\boldsymbol{r};-\boldsymbol{\hat\ell})\cdot\boldsymbol{f}(\boldsymbol{r})=\nonumber\\
\oint_{\mathcal{S}_\mathrm{p}}\mathrm{d}S\, \boldsymbol{v}(\boldsymbol{r};\boldsymbol{\hat\ell})\cdot\boldsymbol{\sigma}_\mathrm{H}^{(0)}(\boldsymbol{r};-\boldsymbol{\hat\ell})\cdot\boldsymbol{\hat n}(\boldsymbol{r})+\int_{\mathcal{V}_\mathrm{f}}\mathrm{d}V\, \boldsymbol{v}(\boldsymbol{r};\boldsymbol{\hat\ell})\cdot\boldsymbol{f}^{(0)}(\boldsymbol{r}). \label{eq:Lorentz}
\end{gather}
Due to the nature of odd viscosity, the auxiliary flow is to be evaluated at the time-reversed spin-momentum density $-\boldsymbol{\hat\ell}$, as indicated by the second argument in 
the
relevant quantities. The flow of interest with $\boldsymbol{f}(\boldsymbol{r})=\rho(\boldsymbol{r})\boldsymbol{E}(\boldsymbol{r})$ satisfies boundary conditions Eqs. \eqref{eq:bc1} and \eqref{eq:bc2}. For the auxiliary flow, we consider an uncharged particle suspended in an odd fluid with no free ions in the solution, $n_i(\boldsymbol{r})=0$ for $i=1,...,N$, and thus $\boldsymbol{f}^{(0)}(\boldsymbol{r})=\boldsymbol{0}$.
In the derivation of \citet{Teubner:1982}, the same stick boundary conditions for the auxiliary flow are taken on $\mathcal{S}_\mathrm{p}$ as the flow of interest. However, for an odd fluid it is essential to set $\boldsymbol{v}^{(0)}(\boldsymbol{r})|_{\boldsymbol{r}\in\mathcal{S}_\mathrm{p}}=\boldsymbol{U}^{(0)}+\boldsymbol{\Omega}^{(0)}\times\boldsymbol{r}$, with $\boldsymbol{U}\neq\boldsymbol{U}^{(0)}$ and $\boldsymbol{\Omega}\neq\boldsymbol{\Omega}^{(0)}$, because the friction and mobility tensors contain antisymmetric contributions. Thus, Eq. \eqref{eq:Lorentz} becomes
\begin{equation}
\boldsymbol{U}^{(0)}\cdot\boldsymbol{F}_\mathrm{H}-\boldsymbol{U}\cdot\boldsymbol{F}_\mathrm{H}^{(0)}+\boldsymbol{\Omega
}^{(0)}\cdot\boldsymbol{T}_\mathrm{H}-\boldsymbol{\Omega}\cdot\boldsymbol{T}_\mathrm{H}^{(0)}=-\int_{\mathcal{V}_\mathrm{f}}\mathrm{d}V\, \rho(\boldsymbol{r})\boldsymbol{v}^{(0)}(\boldsymbol{r};-\boldsymbol{\hat\ell})\cdot\boldsymbol{E}(\boldsymbol{r}).
\end{equation}
We add to this equation $\boldsymbol{U}^{(0)}\cdot\boldsymbol{F}_\mathrm{E}+\boldsymbol{\Omega}^{(0)}\cdot\boldsymbol{T}_\mathrm{E}$, and use that the fluid is force- and torque-free. Furthermore, by introducing the grand friction and mobility matrices
\begin{equation}
\begin{pmatrix}
\boldsymbol{F}_\mathrm{H}^{(0)}\\
\boldsymbol{T}_\mathrm{H}^{(0)}
\end{pmatrix}=
\begin{pmatrix}
\boldsymbol{\zeta}^\mathrm{tt}(\boldsymbol{\hat\ell}) & \boldsymbol{\zeta}^\mathrm{tr}(\boldsymbol{\hat\ell})\\
\boldsymbol{\zeta}^\mathrm{rt}(\boldsymbol{\hat\ell}) & \boldsymbol{\zeta}^\mathrm{rr}(\boldsymbol{\hat\ell})
\end{pmatrix}
\begin{pmatrix}
\boldsymbol{U}^{(0)} \\ \boldsymbol{\Omega}^{(0)}
\end{pmatrix},
\quad 
\begin{pmatrix}
\boldsymbol{\mu}^\mathrm{tt}(\boldsymbol{\hat\ell}) & \boldsymbol{\mu}^\mathrm{tr}(\boldsymbol{\hat\ell})\\
\boldsymbol{\mu}^\mathrm{rt}(\boldsymbol{\hat\ell}) & \boldsymbol{\mu}^\mathrm{rr}(\boldsymbol{\hat\ell})
\end{pmatrix}
=\begin{pmatrix}
\boldsymbol{\zeta}^\mathrm{tt}(\boldsymbol{\hat\ell}) & \boldsymbol{\zeta}^\mathrm{tr}(\boldsymbol{\hat\ell})\\
\boldsymbol{\zeta}^\mathrm{rt}(\boldsymbol{\hat\ell}) & \boldsymbol{\zeta}^\mathrm{rr}(\boldsymbol{\hat\ell})
\end{pmatrix}^{-1} \, ,
\end{equation} 
we find
\begin{gather}
\boldsymbol{\zeta}^\mathrm{tt}(\boldsymbol{\hat\ell})\cdot\boldsymbol{U}+\boldsymbol{\zeta}^\mathrm{tr}(\boldsymbol{\hat\ell})\cdot\boldsymbol{\Omega}=\int_{\mathcal{V}_\mathrm{f}}\mathrm{d}V\, \rho(\boldsymbol{r})\boldsymbol{E}(\boldsymbol{r})\cdot[\mathsfbi{V}(\boldsymbol{r};-\boldsymbol{\hat\ell})-\mathsfbi{I}], \label{eq:bla1}\\
\boldsymbol{\zeta}^\mathrm{rt}(\boldsymbol{\hat\ell})\cdot\boldsymbol{U}+\boldsymbol{\zeta}^\mathrm{rr}(\boldsymbol{\hat\ell})\cdot\boldsymbol{\Omega}=\int_{\mathcal{V}_\mathrm{f}}\mathrm{d}V\, \rho(\boldsymbol{r})\boldsymbol{E}(\boldsymbol{r})\cdot[\mathsfbi{W}(\boldsymbol{r};-\boldsymbol{\hat\ell})-\boldsymbol{r}\cdot\boldsymbol{\epsilon}], \label{eq:bla2}
\end{gather}
where we used the linearity of the equations describing the uncharged auxiliary fluid, following the convention of \citet{HappelBrenner}, to define
\begin{equation}
\boldsymbol{v}^{(0)}(\boldsymbol{r};\boldsymbol{\hat\ell})=\mathsfbi{V}(\boldsymbol{r};\boldsymbol{\hat\ell})\cdot\boldsymbol{U}^{(0)}+\mathsfbi{W}(\boldsymbol{r};\boldsymbol{\hat\ell})\cdot\boldsymbol{\Omega}^{(0)}.
\end{equation}
Furthermore, we used the symmetry property 
$\zeta_{\alpha\beta}^{ab}(\boldsymbol{\hat\ell})=\zeta_{\beta\alpha}^{ba}(-\boldsymbol{\hat\ell})$
for $a,b\in\{\mathrm{t},\mathrm{r}\}$ \citep{Everts:2024}.
Next, we expand the electrostatic potential $\psi(\boldsymbol{r})$, to linear order in $\boldsymbol{E}_\mathrm{ext}$, as
\begin{equation}
\psi(\boldsymbol{r})=\psi^\mathrm{eq}( \boldsymbol{r})+\boldsymbol{\varphi}\cdot\boldsymbol{E}_\mathrm{ext}+... \, , \label{eq:exoo}
\end{equation}
where $\psi^\mathrm{eq}(\boldsymbol{r})$ is the equilibrium electrostatic potential (determined by $\boldsymbol{J}_i^\mathrm{eq}(\boldsymbol{r})=\boldsymbol{0}$, $\boldsymbol{v}^\mathrm{eq}(\boldsymbol{r})=\boldsymbol{0}$). For the equilibrium situation, we consider particles with
fixed zeta potential -- the electrostatic potential at the slipping plane --, i.e.
 $\psi^\mathrm{eq}(\boldsymbol{r})|_{\boldsymbol{r}\in\mathcal{S}_\mathrm{p}}=\Psi_0$. Using Eq. \eqref{eq:exoo}, we find the following expansion for the electric body force: 
\begin{equation}
\rho(\boldsymbol{r})\boldsymbol{E}(\boldsymbol{r})=\varepsilon\nabla^2\psi^\mathrm{eq}(\boldsymbol{r})\nabla\psi^\mathrm{eq}(\boldsymbol{r})+\varepsilon[\nabla^2\psi^\mathrm{eq}(\boldsymbol{r})\nabla\boldsymbol{\varphi}(\boldsymbol{r})+\nabla\psi^\mathrm{eq}(\boldsymbol{r})\nabla^2\boldsymbol{\varphi}(\boldsymbol{r})]\cdot\boldsymbol{E}_{\mathrm{ext}}+... \, .
\end{equation}
When inserted into Eqs. \eqref{eq:bla1} and \eqref{eq:bla2}, the first term on the right-hand side vanishes. 
At low zeta potentials $|e\Psi_0/(k_\mathrm{B}T)|\ll1$, (the Henry approximation) and/or thin electric double layers (Smoluchowski limit), the other terms can be simplified.
In this case,
\begin{equation}
\nabla^2\boldsymbol{\varphi}(\boldsymbol{r})=\boldsymbol{0}, \quad \boldsymbol{\hat n}(\boldsymbol{r})\cdot\nabla\boldsymbol{\varphi}(\boldsymbol{r})=\boldsymbol{0}, \quad \boldsymbol{\varphi}(\boldsymbol{r})\rightarrow -\boldsymbol{r}+O(1/r^2), \quad r\rightarrow\infty.
\end{equation}
Eqs. \eqref{eq:bla1} and \eqref{eq:bla2} simplify to
\begin{gather}
\boldsymbol{\zeta}^\mathrm{tt}(\boldsymbol{\hat\ell})\cdot\boldsymbol{U}+\boldsymbol{\zeta}^\mathrm{tr}(\boldsymbol{\hat\ell})\cdot\boldsymbol{\Omega}=\left\{\varepsilon\int_{\mathcal{V}_\mathrm{f}}\mathrm{d}V\, \nabla^2\psi^\mathrm{eq}(\boldsymbol{r})  [\mathsfbi{V}^\top(\boldsymbol{r};-\boldsymbol{\hat\ell})-\mathsfbi{I}]\cdot\nabla\boldsymbol{\varphi}(\boldsymbol{r})\right\}\cdot\boldsymbol{E}_\mathrm{ext}, \label{eq:inter}\\
\boldsymbol{\zeta}^\mathrm{rt}(\boldsymbol{\hat\ell})\cdot\boldsymbol{U}+\boldsymbol{\zeta}^\mathrm{rr}(\boldsymbol{\hat\ell})\cdot\boldsymbol{\Omega}=\left\{\varepsilon\int_{\mathcal{V}_\mathrm{f}}\mathrm{d}V\, \nabla^2\psi^\mathrm{eq}(\boldsymbol{r})[\mathsfbi{W}^\top(\boldsymbol{r};-\boldsymbol{\hat\ell})+\boldsymbol{r}\cdot\boldsymbol{\epsilon}] \cdot\nabla\boldsymbol{\varphi}(\boldsymbol{r})\right\}\cdot\boldsymbol{E}_\mathrm{ext}\nonumber,
\end{gather}
with $\top$ denoting the tensor transpose.
Eq. \eqref{eq:inter} can be inverted to obtain the electrophoretic mobility and electrorotation tensor as defined in Eq. \eqref{eq:eletensors}. Furthermore, note that Eq. \eqref{eq:inter} does not depend on $\mathsfbi{D}_i(\boldsymbol{\hat\ell})$.

\section{Electrophoretic mobility for a sphere with uniform zeta potential}
An uncharged sphere of radius $a$ in an odd fluid described by Eq. \eqref{eq:hyd} exhibits no translational-rotational coupling \citep{Everts:2024}. In this case, combining Eq.~\eqref{eq:eletensors} with Eq.~\eqref{eq:inter} gives for the electrophoretic mobilities
\begin{gather}
\boldsymbol{\mu}^\mathrm{tE}(\boldsymbol{\hat\ell})=\boldsymbol{\mu}^\mathrm{tt}(\boldsymbol{\hat\ell})\cdot\left\{\varepsilon\int_{\mathcal{V}_\mathrm{f}}\mathrm{d}V\, \nabla^2\psi^\mathrm{eq}(\boldsymbol{r})[\mathsfbi{V}^\top(\boldsymbol{r};-\boldsymbol{\hat\ell})-\mathsfbi{I}]\cdot\nabla\boldsymbol{\varphi}(\boldsymbol{r})\right\}, \label{eq:elemob}  \\
\boldsymbol{\mu}^\mathrm{rE}(\boldsymbol{\hat\ell})=\boldsymbol{\mu}^\mathrm{rr}(\boldsymbol{\hat\ell})\cdot\left\{\varepsilon\int_{\mathcal{V}_\mathrm{f}}\mathrm{d}V\, \nabla^2\psi^\mathrm{eq}(\boldsymbol{r})[\mathsfbi{W}^\top(\boldsymbol{r};-\boldsymbol{\hat\ell})+\boldsymbol{r}\cdot\boldsymbol{\epsilon}] \cdot\nabla\boldsymbol{\varphi}(\boldsymbol{r})\right\}. \label{eq:elemob2}
\end{gather}
Note that all quantities given in  
Eqs. \eqref{eq:elemob} and \eqref{eq:elemob2}
are analytically known. The electrostatic quantities are
\begin{equation}
\psi^\mathrm{eq}(\boldsymbol{r})=\Psi_0\frac{a}{r}\mathrm{e}^{\kappa(a-r)}, \quad \boldsymbol{\varphi}(\boldsymbol{r})=-\left(1+\frac{a^3}{2r^3}\right)\boldsymbol{r}, \label{eq:electrostat}
\end{equation}
where $\kappa^{-1}$ is the Debye screening length with $\kappa^2=e^2/(\varepsilon k_\mathrm{B}T)\sum_{i=1}^N z_i^2 n_i^\infty$.
For electrorotation,
$\mathsfbi{W}(\boldsymbol{r};\boldsymbol{\hat\ell})=(a^3/r^3)(\boldsymbol{r}\cdot\boldsymbol{\epsilon})$ \citep{Hosaka:2024}, 
thus it follows that 
${\mu}^\mathrm{rE}_{\alpha\beta}(\boldsymbol{\hat\ell})=0$ for all $\alpha,\beta\in\{x,y,z\}$. For electrophoresis, the  tensor $\mathsfbi{V}(\boldsymbol{r};\boldsymbol{\hat\ell})$ is analytically known in 
closed form \citep{Meissner:2025}, and can be expressed as 
\begin{equation}
\mathsfbi{V}(\boldsymbol{r};\boldsymbol{\hat\ell})=\mathcal{L}_0\mathsfbi{G}(\boldsymbol{r};\boldsymbol{\hat\ell})\cdot\boldsymbol{\zeta}^\mathrm{tt}(\boldsymbol{\hat\ell}), 
\quad 
\mathcal{L}_0 = \sum_{{k}=0}^\infty\frac{a^{2{k}}}{(2{k}+1)!}(\nabla^2)^{{k}}, \label{eq:sing}
\end{equation}
with $\mathsfbi{G}(\boldsymbol{r};\boldsymbol{\hat\ell})$ the odd Oseen tensor \citep{Everts:2024}. 
Although it is known analytically, we do not need its explicit form. The only relevant property is that it can be expressed as $\mathsfbi{G}(\boldsymbol{r};\boldsymbol{\hat\ell})=(a/r)\mathsfbi{g}(\boldsymbol{\hat r};\boldsymbol{\hat\ell})$, see Appendix \ref{app}. 
The final quantity that we need 
is 
\begin{equation}
\boldsymbol{\mu}^\mathrm{tt}(\boldsymbol{\hat\ell})=
\frac{1}{6\pi\eta_\mathrm{s}a} \left[m_\parallel(\gamma)\boldsymbol{\hat\ell}\boldsymbol{\hat\ell}+m_\perp(\gamma)(\mathsfbi{I}-\boldsymbol{\hat\ell}\boldsymbol{\hat\ell})+m_\mathrm{o}(\gamma)(\boldsymbol{\epsilon}\cdot\boldsymbol{\hat\ell})\right]
  . \label{eq:mutt}
\end{equation}
The closed-form analytical expressions for 
$m_\parallel(\gamma)$, $m_\perp(\gamma)$ and $m_\mathrm{o}(\gamma)$
as functions of arbitrary values of $\gamma=\eta_\mathrm{o}/\eta_\mathrm{s}$ can be found in \cite{Everts:2024}. For $\gamma\rightarrow 0$, we have 
$m_\parallel(\gamma)\rightarrow 1$, $m_\perp(\gamma)\rightarrow 1$,
and $m_\mathrm{o}(\gamma)\rightarrow 0$, retrieving the Newtonian-fluid result for the translational mobility.

\subsection{Hückel limit}
First, we consider the limit $\kappa a\ll 1$. This is the case of an infinite electric double layer and has been first analysed for simple fluids by \citet{Huckel}.  We are considering stationary motion, so there is a force balance between the electric force and the hydrodynamic drag force,
$\boldsymbol{\zeta}^{\mathrm{tt}}(\hat{\boldsymbol{\ell}})\cdot\boldsymbol{U}=Ze\boldsymbol{E}_\mathrm{ext}$, or $\boldsymbol{U}=Ze\boldsymbol{\mu}^{\mathrm{tt}}(\hat{\boldsymbol{\ell}})\cdot\boldsymbol{E}_\mathrm{ext}, $
with $Ze$ the total charge of the particle.
In equilibrium, with constant-potential boundary conditions on $\mathcal{S}_\mathrm{p}$, we have $Ze=4\pi\varepsilon a\Psi_0$ for $\kappa a\ll 1$. We conclude that the electrophoretic mobility is $\boldsymbol{\mu}^\mathrm{tE}(\hat{\boldsymbol{\ell}})=4\pi\varepsilon a\Psi_0\boldsymbol{\mu}^{\mathrm{tt}}(\hat{\boldsymbol{\ell}})$ with components
\begin{align}
\boldsymbol{\mu}^\mathrm{tE}(\hat{\boldsymbol{\ell}})=\frac{2\varepsilon}{3\eta_\mathrm{s}}\Psi_0\left[m_\parallel(\gamma)\boldsymbol{\hat\ell}\boldsymbol{\hat\ell}+m_\perp(\gamma)(\mathsfbi{I}-\boldsymbol{\hat\ell}\boldsymbol{\hat\ell})+m_\mathrm{o}(\gamma)(\boldsymbol{\epsilon}\cdot\boldsymbol{\hat\ell})\right], \quad \kappa a \ll 1. \label{eq:Huckel}
\end{align}
Equivalently, one can neglect $\mathsfbi{V}(\boldsymbol{r};\boldsymbol{\hat\ell})$ in Eq. \eqref{eq:elemob} \citep{Teubner:1982} and by global charge neutrality one would find the same result. 
Note that for $\gamma\rightarrow 0$, we retrieve the Newtonian result for the H\"uckel limit.
Moreover, the electrophoretic mobility tensor contains distinct contributions in the parallel, perpendicular and axial (i.e., odd) components, in contrast to the case of a homogeneously charged spheres suspended in a Newtonian fluid where $\boldsymbol{\mu}^\mathrm{tE}(\hat{\boldsymbol{\ell}})$ is isotropic.

\subsection{Smoluchowski limit}
For thin electric double layers, $\kappa a \gg 1$, first analysed for Newtonian fluids by \cite{Smoluchowski}, we 
transform Eq. \eqref{eq:elemob} by two partial integrations and applying the boundary condition for $\mathsfbi{V}(\boldsymbol{r};\boldsymbol{\hat\ell})|_{\boldsymbol{r}\in\mathcal{S}_\mathrm{p}}=\mathsfbi{I}$. The volume term can be neglected in this large-screening limit and we find
\begin{equation}
\boldsymbol{\mu}^\mathrm{tE}({\boldsymbol{\hat\ell}})=-\varepsilon\Psi_0\int_{\mathcal{S}_\mathrm{p}}dS\, \boldsymbol{\mu}^\mathrm{tt}(\hat{\boldsymbol{\ell}})\cdot\left\{[\boldsymbol{\hat n}(\boldsymbol{r})\cdot\nabla]\mathsfbi{V}^\top(\boldsymbol{r};-{\boldsymbol{\hat \ell}})\right\}\cdot\nabla\boldsymbol{\varphi}(\boldsymbol{r}). \label{eq:Cartman}
\end{equation}
For a sphere $\boldsymbol{\hat n}=-\boldsymbol{\hat r}$ and  
after
using Eq. \eqref{eq:sing} we find that Eq. \eqref{eq:Cartman} simplifies to
\begin{equation}
\boldsymbol{\mu}^\mathrm{tE}({\boldsymbol{\hat\ell}})=-6\pi\varepsilon a^2\Psi_0\left[\frac{\partial}{\partial r}\overline{\mathcal{L}_0\mathsfbi{G}(\boldsymbol{r};\boldsymbol{\hat\ell})\cdot(\mathsfbi{I}-\boldsymbol{\hat r}\boldsymbol{\hat r})}\right]_{r=a}. \label{eq:haha}
\end{equation}
Here, we denote a surface average over the unit sphere $S^2$ as $\overline{(...)}=(4\pi)^{-1}\int_{S^2}\mathrm{d}^2\boldsymbol{\hat r}\, (...)$.
Now, for fixed $\alpha$, we may view $\mathcal{L}_0G_{\alpha\beta}(\boldsymbol{r};\boldsymbol{\hat \ell})$ as an incompressible vector field that is constant on $\mathcal{S}_\mathrm{p}$. 
Therefore, its normal component has {a} vanishing gradient on $\mathcal{S}_\mathrm{p}$, which means that $\frac{\partial}{\partial r}\overline{\mathcal{L}_0\mathsfbi{G}(\boldsymbol{r};\boldsymbol{\hat\ell})\cdot\boldsymbol{\hat r}\boldsymbol{\hat r}}$
vanishes on $\mathcal{S}_\mathrm{p}$. The remaining part equals $-1/a$ upon using 
Eq. \eqref{eq:Bogdan1}; see Appendix \ref{app}. 
We find $\boldsymbol{\mu}^\mathrm{tE}({\boldsymbol{\hat\ell}})=6\pi\varepsilon a\Psi_0\boldsymbol{\mu}^\mathrm{tt}(\boldsymbol{\hat\ell})$ with components
\begin{equation}
\boldsymbol{\mu}^\mathrm{tE}({\boldsymbol{\hat\ell}})
=\frac{\varepsilon}{\eta_\mathrm{s}}\Psi_0\left[m_\parallel(\gamma)\boldsymbol{\hat\ell}\boldsymbol{\hat\ell}+m_\perp(\gamma)(\mathsfbi{I}-\boldsymbol{\hat\ell}\boldsymbol{\hat\ell})+m_\mathrm{o}(\gamma)(\boldsymbol{\epsilon}\cdot\boldsymbol{\hat\ell})\right]
, \quad \kappa a\gg 1. \label{eq:Smo}
\end{equation}
Note that this result is not necessarily restricted to the case $|e\Psi_0/(k_\mathrm{B}T)|\ll1$, since it also applies to particles with a local radius of curvature much larger than $\kappa^{-1}$. Furthermore, we retrieve the Newtonian result for the Smoluchowski limit when $\gamma\rightarrow 0$. Interestingly, the fluid anisotropy still persists in $\boldsymbol{\mu}^\mathrm{tE}({\boldsymbol{\hat\ell}})$ even for short Debye screening lengths. This is in sharp contrast to charged anisotropic particles in isotropic Newtonian fluids, where the electrophoretic mobility is isotropic and independent of particle shape under the same screening conditions.

\subsection{Henry approximation}
Next, we analyse electrophoresis for intermediate Debye screening length
and small zeta 
potentials.
Combining Eqs. \eqref{eq:elemob} and \eqref{eq:sing} gives
\begin{equation}
\boldsymbol{\mu}^\mathrm{tE}(\hat{\boldsymbol{\ell}})=\varepsilon\int_{\mathcal{V}_\mathrm{f}}\mathrm{d}V\, \nabla^2\psi^\mathrm{eq}(r)[\mathcal{L}_0\mathsfbi{G}({\bf r};\hat{\boldsymbol{\ell}})-\boldsymbol{\mu}^\mathrm{tt}(\hat{\boldsymbol{\ell}})]\cdot\nabla\boldsymbol{\varphi}({\bf r}). \label{eq:pfff}
\end{equation}
We write $\int_{\mathcal{V}_\mathrm{f}}\mathrm{d}V=\int_a^\infty\mathrm{d}r\, r^2\int_{S^2}\mathrm{d}^2\boldsymbol{\hat r}$, and we obtain
\begin{equation}
\boldsymbol{\mu}^\mathrm{tE}(\hat{\boldsymbol{\ell}})=4\pi\varepsilon\kappa^2a\Psi_0\int_a^\infty dr\, r\mathrm{e}^{\kappa(a-r)}\left\{\,\overline{[\mathcal{L}_0\mathsfbi{G}({\bf r};\hat{\boldsymbol{\ell}})]\cdot\nabla\boldsymbol{\varphi}({\bf r})}+\boldsymbol{\mu}^\mathrm{tt}(\hat{\boldsymbol{\ell}})\right\}, \label{eq:Reinier}
\end{equation}
where we used that $\overline{\nabla\boldsymbol{\varphi}(\boldsymbol{r})}=-\mathsfbi{I}$.
After applying the results
from
Eqs. \eqref{eq:Bogdan1} and \eqref{eq:Bogdan2} (see Appendix \ref{app}), we find
\begin{equation}
\overline{\mathcal{L}_0\mathsfbi{G}(\boldsymbol{r};\hat{\boldsymbol{\ell}})\cdot\nabla\boldsymbol{\varphi}({\bf r})}=\left(-\frac{a}{r}+\frac{a^4}{4r^4}-\frac{a^6}{4r^6}\right)\boldsymbol{\mu}^\mathrm{tt}(\hat{\boldsymbol{\ell}}). \label{eq:essential}
\end{equation}
We conclude that $\boldsymbol{\mu}^\mathrm{tE}(\hat{\boldsymbol{\ell}})=6\pi\varepsilon a\Psi_0f(\kappa a)\boldsymbol{\mu}^\mathrm{tt}(\hat{\boldsymbol{\ell}})$, with components
\begin{gather}
\boldsymbol{\mu}^\mathrm{tE}(\hat{\boldsymbol{\ell}})=\frac{\varepsilon}{\eta_\mathrm{s}}f(\kappa a)\Psi_0\left[m_\parallel(\gamma)\boldsymbol{\hat\ell}\boldsymbol{\hat\ell}+m_\perp(\gamma)(\mathsfbi{I}-\boldsymbol{\hat\ell}\boldsymbol{\hat\ell})+m_\mathrm{o}(\gamma)(\boldsymbol{\epsilon}\cdot\boldsymbol{\hat\ell})\right]
, \label{eq:Reinier2}
\end{gather}
and the Henry function $f$ \citep{Henry:1931} defined by
\begin{equation}
f(x)=\frac{2}{3}x^2\mathrm{e}^x\int_1^\infty dt\, t\mathrm{e}^{-xt}\left(1-\frac{1}{t}+\frac{1}{4t^4}-\frac{1}{4t^6}\right)=1-\mathrm{e}^x[5E_7(x)-2E_5(x)],
\end{equation}
for $x>0$.
Here, $E_{{k}}(x)=x^{{k}-1}\int_x^\infty dt\, {\mathrm{e}^{-t}}/{t^{{k}}}$ are exponentials integrals. The function $f$ has the properties $\lim_{x\rightarrow 0}f(x)=2/3$ and  $\lim_{x\rightarrow \infty}f(x)=1$. 
Therefore, Eq. \eqref{eq:Reinier2} reproduces both the Hückel limit (Eq.~\eqref{eq:Huckel}), for $\kappa a\ll 1$, and the Smoluchowski limit (Eq.~\eqref{eq:Smo}), for $\kappa a\gg 1$. 
The result for a Newtonian fluid is retrieved for $\gamma\rightarrow 0$ as well. 
Again, the electrophoretic mobility tensor contains distinct parallel, perpendicular, and axial components, while the modification by the Henry function is identical to that in isotropic Newtonian fluids. We note that for $\gamma\ll1$, $m_{\perp,\parallel}(\gamma)=O(\gamma^2)$, while $m_\mathrm{o}(\gamma)=\gamma/2+O(\gamma^3)$ \citep{Everts:2024}. In this regime, the linear growth of the axial component of $\boldsymbol{\mu}^{\mathrm{tE}}(\boldsymbol{\hat\ell})$ with $\gamma$, which is proportional to the steady-state intrinsic rotation-rate vector of the fluid particles \citep{Markovich:2021}, is an important experimental signature of odd electrophoresis.

We have also obtained Eq. \eqref{eq:Reinier2} from a numerical computation using Eqs. \eqref{eq:elemob}, \eqref{eq:electrostat} and \eqref{eq:mutt}, with the exact closed-form solution for $\mathsfbi{V}(\boldsymbol{r};\hat{\boldsymbol{\ell}})$ listed in \cite{Meissner:2025}. Considering the complicated analytical form of this quantity, it shows the power of the singularity representation Eq. \eqref{eq:sing}: it showcases how numerics can be completely circumvented due to the uncharged translating sphere having constant tractions on its surface.
By the same properties, we find the remarkable result that $\boldsymbol{\mu}^\mathrm{tE}(\boldsymbol{\hat\ell})$ is proportional to $\boldsymbol{\mu}^\mathrm{tt}(\boldsymbol{\hat\ell})$ for a charged sphere within the Henry approximation and/or Smoluchowski limit.

\section{Conclusions and outlook}
Our paper provides an exact analytical solution for the electrophoretic mobility of a charged spherical particle suspended in an electrolyte solution with odd viscosity.
We have introduced, for the first time, the concept of a charged chiral active fluid and generalised the expressions for the electrophoresis of arbitrarily shaped particles dissolved in an odd fluid. 
We explicitly calculated the electrophoretic mobility tensor for spherical particles dissolved in fluids with odd viscosity for low zeta potentials, small external electric fields, and for general Debye screening length. Moreover, our results are valid for arbitrary values of the odd-viscosity coefficient. As in isotropic Newtonian fluids, we find no electrorotation for a homogeneously charged sphere, and find that the (translational) electrophoretic mobility of a sphere is proportional to the translational mobility of an uncharged sphere multiplied by the Henry function.  Furthermore, we provide closed-form analytical expressions for both the Hückel and Smoluchowski limits.
Our work demonstrates that odd viscosity leads to directional asymmetries in the electrophoretic mobility tensor, suggesting mechanisms for active control of charged colloidal motion in systems where odd viscosity is prevalent.
Furthermore, these anisotropies are still present in the Smoluchowski limit.
Thus, our work bridges the gap between classical electrophoresis and odd fluids.

In our theoretical framework, we focused on the regime of linear electrophoresis, where we expanded around equilibrium for small external electric fields. Subsequently, we used a low zeta-potential approximation. Furthermore, we have assumed that the dielectric properties are isotropic. Relaxing these assumptions may 
lead 
to effects that are absent from the present analysis. For example, it is well known that a non-uniform equilibrium zeta potential can cause electrorotation of a charged particle in an isotropic Newtonian fluid -- even within the Smoluchowski limit \citep{Teubner:1982}. It would be interesting to see whether this effect can also occur for a particle with uniform zeta potential in an anisotropic dielectric medium, where the anisotropy originates from a non-vanishing spin-momentum of the fluid. 
In general,
dielectric anisotropy 
causes
the surface charge to be inhomogeneous for a constant-potential particle, as one can deduce from equilibrium calculations of charged (colloidal) spheres in ion-doped nematic fluids \citep{Everts:2021}.  We expect that an odd (axial) dielectric contribution could generate analogous effects, leading to electrorotation through the coupling of induced surface-charge multipoles to an external electric field.

Alternatively, at high zeta potentials, $|e\Psi_0/(k_\mathrm{B}T)| \gg 1$, we hypothesise that similar effects can occur through significant surface conduction and polarisation effects \citep{Dukhin:1993},  which
are not included in our present work. Other possible directions for future work include the extension of our analysis to multiple charged particles to assess the role of (odd) hydrodynamic interactions \citep{yuan2023} on odd electrophoresis. It would be interesting to see how many-body effects will alter the results presented in this work and how they differ from the isotropic Newtonian case \citep{Chen:1988}.

\begin{bmhead}[Acknowledgements.]
We thank Jerzy Gamdzyk for insightful discussions.
\end{bmhead}

\begin{bmhead}[Funding.]
We acknowledge funding from the National Science Centre, Poland, within the OPUS LAP grant no. 2024/55/I/ST3/00998.
\end{bmhead}

\begin{bmhead}[Declaration of interests.]
The authors report no conflict of interest.
\end{bmhead}

\begin{appen}
\section{}\label{app}
In this Appendix, we explicitly compute $\overline{\mathcal{L}_0\mathsfbi{G}(\boldsymbol{r};\boldsymbol{\hat\ell})}$ and $\overline{\mathcal{L}_0\mathsfbi{G}(\boldsymbol{r};\boldsymbol{\hat\ell})\cdot\boldsymbol{\hat r}\boldsymbol{\hat r}}$, which were necessary to compute Eqs. \eqref{eq:haha} and \eqref{eq:essential}.  Our starting point is that for stick boundary conditions, we have the single-layer integral representation theorem
\begin{equation}
\boldsymbol{u}(\boldsymbol{r};\boldsymbol{\hat\ell})=\int_{\mathcal{S}_\mathrm{p}}\mathrm{d}S\, \mathsfbi{G}(\boldsymbol{r}-\boldsymbol{r}';\boldsymbol{\hat\ell})\cdot[\boldsymbol{\sigma}_\mathrm{H}(\boldsymbol{r}')\cdot\boldsymbol{\hat n}(\boldsymbol{r}')], \quad \boldsymbol{u}(\boldsymbol{r};\boldsymbol{\hat\ell})=\begin{cases}
\boldsymbol{U}, \quad \boldsymbol{r}\in\mathcal{V}_\mathrm{p},\\
\boldsymbol{v}(\boldsymbol{r};\boldsymbol{\hat\ell}),\quad \boldsymbol{r}\in\mathcal{V}_\mathrm{f}, \label{eq:intrep}
\end{cases}
\end{equation}
where we only need the result for a translating sphere. In this case the tractions $\boldsymbol{\sigma}_\mathrm{H}(\boldsymbol{r}')\cdot\boldsymbol{\hat n}(\boldsymbol{r}')$ are constant on $\mathcal{S}_\mathrm{p}$ and equal to $\boldsymbol{F}_\mathrm{H}/(4\pi a^2)$ as was proved by  \cite{Everts:2024}. Noting that the Greens function can be written as $\mathsfbi{G}(\boldsymbol{r};\boldsymbol{\hat\ell})=(a/r)\mathsfbi{g}(\boldsymbol{\hat r};\boldsymbol{\hat\ell})$ \citep{Everts:2024} and evaluating Eq. \eqref{eq:intrep} for $\boldsymbol{r}=\boldsymbol{0}$, we find that $\boldsymbol{U}=\overline{\mathsfbi{g}(\boldsymbol{\hat r};\boldsymbol{\hat\ell})}\cdot\boldsymbol{F}_\mathrm{H}$. We conclude that $\overline{\mathsfbi{g}(\boldsymbol{\hat r};\boldsymbol{\hat\ell})}=\boldsymbol{\mu}^\mathrm{tt}(\boldsymbol{\hat\ell})$. Similar results are also noted by \cite{Khain:2024}. We now proceed to the computation of the relevant quantities.

\subsection{Computation of the surface average of $\mathcal{L}_0\mathsfbi{G}$ over $S^2$.}

\label{sec:subad}
First, we show that $\overline{(\nabla^2)^{{k}}\mathsfbi{G}(\boldsymbol{r};\boldsymbol{\hat\ell})}=0$ for ${{k}}\geq1$. We adopt spherical coordinates where $\boldsymbol{\hat\ell}$ is parallel to the $z$ axis. In these coordinates, we decompose the Laplacian as $\nabla^2=\nabla_r^2+(1/r^2)\Delta_{S^2}$, where $\nabla_r^2(...)=(1/r^2)\partial_r^2[r^2(...)]$, and $\Delta_{S^2}$ is the Laplace-Beltrami operator on the unit sphere $S^2$ \citep{Flanders}. It follows by direct computation that $\nabla^2\mathsfbi{G}(\boldsymbol{r};\boldsymbol{\hat\ell})=(a/r^3)\Delta_{S^2}\mathsfbi{g}(\boldsymbol{\hat r};\boldsymbol{\hat\ell})$ for $r>0$. The key step to proceed further, is to observe that  $\overline{\Delta_{S^2}\mathsfbi{g}(\boldsymbol{\hat r};\boldsymbol{\hat\ell})}=0$, which can be shown by application of the divergence theorem on a curved manifold and noting that $S^2$ is a compact manifold without boundary. We conclude that $\overline{\nabla^2\mathsfbi{G}(\boldsymbol{r};\boldsymbol{\hat\ell})}=0$ for $r>0$.
Now suppose that the statement $\overline{(\nabla^2)^{{k}}\mathsfbi{G}(\boldsymbol{r};\boldsymbol{\hat\ell})}=0$ is true for ${{k}}={l}$. Then $\overline{(\nabla^2)^{{l}+1}\mathsfbi{G}(\boldsymbol{r};\boldsymbol{\hat\ell})}=\nabla_r^2[\overline{(\nabla^2)^{{l}}\mathsfbi{G}(\boldsymbol{r};\boldsymbol{\hat\ell})}]+(1/r^2)\overline{\Delta_{S^2}[(\nabla^2)^{{l}}\mathsfbi{G}(\boldsymbol{r};\boldsymbol{\hat\ell})}]$. The first term vanishes because of the induction hypothesis. The second term vanishes again by virtue of the divergence theorem on $S^2$. This completes the proof of our initial statement. Now, using the linearity of the surface average, we obtain
\begin{equation}
\overline{\mathcal{L}_0\mathsfbi{G}(\boldsymbol{r};\boldsymbol{\hat\ell})}=\overline{\mathsfbi{G}(\boldsymbol{r};\boldsymbol{\hat\ell})}=\frac{a}{r}\overline{\mathsfbi{g}(\boldsymbol{\hat r};\boldsymbol{\hat\ell})}=\frac{a}{r}\boldsymbol{\mu}^\mathrm{tt}(\boldsymbol{\hat\ell}) .\label{eq:Bogdan1}
\end{equation}

\subsection{Computation of the surface average of $\mathcal{L}_0\mathsfbi{G} \cdot\boldsymbol{\hat r}\boldsymbol{\hat r}$ over $S^2$.}
First, we show that $\overline{(\nabla^2)^k\mathsfbi{G}(\boldsymbol{r};\boldsymbol{\hat\ell})\cdot\boldsymbol{\hat r}\boldsymbol{\hat r}}=0$ for $k\geq2$.
Because $S^2$ is a compact manifold without boundary, we have the following Greens identity for two tensors $\mathsfbi{A}$ and $\mathsfbi{B}$ defined on $S^2$:
\begin{equation}
\overline{\mathsfbi{A}(\boldsymbol{\hat r})\cdot\Delta_{S^2}\mathsfbi{B}(\boldsymbol{\hat r})}=\overline{[\Delta_{S^2}\mathsfbi{A}(\boldsymbol{\hat r})]\cdot\mathsfbi{B}(\boldsymbol{\hat r})}.
\end{equation}
Now take $\mathsfbi{A}(\boldsymbol{\hat r})=\Delta_{S^2}\mathsfbi{g}(\boldsymbol{\hat r};\boldsymbol{\hat\ell})$ and $\mathsfbi{B}(\boldsymbol{\hat r})=\boldsymbol{\hat r}\boldsymbol{\hat r}$. This results in
\begin{equation}
\overline{[\Delta_{S^2}\Delta_{S^2}\mathsfbi{g}(\boldsymbol{\hat r};\boldsymbol{\hat\ell})]\cdot\boldsymbol{\hat r}\boldsymbol{\hat r}}=-6\overline{[\Delta_{S^2}\mathsfbi{g}(\boldsymbol{\hat r};\boldsymbol{\hat\ell})]\cdot\boldsymbol{\hat r}\boldsymbol{\hat r}}, \label{eq:kas}
\end{equation}
where we used the divergence theorem on $S^2$ and $\Delta_{S^2}\boldsymbol{\hat r}\boldsymbol{\hat r}=2(\mathsfbi{I}-3\boldsymbol{\hat r}\boldsymbol{\hat r})$. Furthermore, by direct computation, we find
\begin{equation}
[\nabla^2\nabla^2\mathsfbi{G}(\boldsymbol{r};\boldsymbol{\hat\ell})]\cdot\boldsymbol{\hat r}\boldsymbol{\hat r}=\frac{6a}{r^5}[\Delta_{S^2}\mathsfbi{g}(\boldsymbol{\hat r};\boldsymbol{\hat\ell})]\cdot\boldsymbol{\hat r}\boldsymbol{\hat r}+\frac{a}{r^5}[\Delta_{S^2}\Delta_{S^2}\mathsfbi{g}(\boldsymbol{\hat r};\boldsymbol{\hat\ell})]\cdot\boldsymbol{\hat r}\boldsymbol{\hat r}.
\end{equation}
By taking the average over $S^2$ of this equation and using Eq. \eqref{eq:kas}, we find $\overline{[\nabla^2\nabla^2\mathsfbi{G}(\boldsymbol{r};\boldsymbol{\hat\ell})]\cdot\boldsymbol{\hat r}\boldsymbol{\hat r}}=0$. The statement then follows by induction using steps similar to those in Sec. \ref{sec:subad}. 

Now we proceed with the computation of the remaining terms. It follows from the above that
\begin{equation}
\overline{[\mathcal{L}_0\mathsfbi{G}(\boldsymbol{r};\boldsymbol{\hat\ell})]\cdot\boldsymbol{\hat r}\boldsymbol{\hat r}}=\overline{\left[\left(1+\frac{a^2}{6}\nabla^2\right)\mathsfbi{G}(\boldsymbol{r};\boldsymbol{\hat\ell})\right]\cdot\boldsymbol{\hat r}\boldsymbol{\hat r}}. \label{eq:syl}
\end{equation}
To obtain the first term in Eq. \eqref{eq:syl}, we find by explicit calculation in spherical coordinates and partial integration that
\begin{equation}
\overline{[\nabla\cdot\mathsfbi{G}^\top(\boldsymbol{r};\boldsymbol{\hat\ell})]\boldsymbol{\hat r}}=-\frac{a}{r^2}\overline{\mathsfbi{g}(\boldsymbol{\hat r};\boldsymbol{\hat\ell})\cdot(\mathsfbi{I}-2\boldsymbol{\hat r}\boldsymbol{\hat r})}, \quad r>0.
\end{equation}
Using the incompressibility condition, it then follows that $\overline{\mathsfbi{g}(\boldsymbol{\hat r};\boldsymbol{\hat\ell})\cdot(\mathsfbi{I}-2\boldsymbol{\hat r}\boldsymbol{\hat r})}=0$. Application  of this result gives
\begin{equation}
\overline{\mathsfbi{G}(\boldsymbol{r};\boldsymbol{\hat\ell})\cdot\boldsymbol{\hat r}\boldsymbol{\hat r}}=\frac{a}{r}\overline{\mathsfbi{g}(\boldsymbol{\hat r};\boldsymbol{\hat\ell})\cdot\boldsymbol{\hat r}\boldsymbol{\hat r}}=\frac{a}{2r}\overline{\mathsfbi{g}(\boldsymbol{\hat r};\boldsymbol{\hat\ell})}=\frac{a}{2r}\boldsymbol{\mu}^\mathrm{tt}(\boldsymbol{\hat\ell}). \label{eq:syl2}
\end{equation}
For the second term in Eq. \eqref{eq:syl}, we need to find $\overline{[\nabla^2\mathsfbi{G}(\boldsymbol{r};\boldsymbol{\hat\ell})]\cdot\boldsymbol{\hat r}\boldsymbol{\hat r}}=(a/r^3)\overline{[\Delta_{S^2}\mathsfbi{g}(\boldsymbol{\hat r};\boldsymbol{\hat\ell})]\cdot\boldsymbol{\hat r}\boldsymbol{\hat r}}$. This contribution is calculated by first evaluating Eq. \eqref{eq:syl} for $r=a$:
\begin{equation}
\overline{[\mathcal{L}_0\mathsfbi{G}(a\boldsymbol{\hat r};\boldsymbol{\hat\ell})]\cdot\boldsymbol{\hat r}\boldsymbol{\hat r}}=\overline{\mathsfbi{g}(\boldsymbol{\hat r};\boldsymbol{\hat\ell})\cdot\boldsymbol{\hat r}\boldsymbol{\hat r}}+\frac{1}{6}\overline{[\Delta_{S^2}\mathsfbi{g}(\boldsymbol{\hat r};\boldsymbol{\hat\ell})]\cdot\boldsymbol{\hat r}\boldsymbol{\hat r}}. \label{eq:syl3}
\end{equation}
However, $\mathcal{L}_0\mathsfbi{G}(a\boldsymbol{\hat r};\boldsymbol{\hat\ell})=\boldsymbol{\mu}^\mathrm{tt}(\boldsymbol{\hat\ell})$ and $\overline{\boldsymbol{\hat r}\boldsymbol{\hat r}}=1/3$. 
We conclude that $\overline{[\Delta_{S^2}\mathsfbi{g}(\boldsymbol{\hat r};\boldsymbol{\hat\ell})]\cdot\boldsymbol{\hat r}\boldsymbol{\hat r}}=-\boldsymbol{\mu}^\mathrm{tt}(\boldsymbol{\hat\ell})$ and therefore $\overline{[\nabla^2\mathsfbi{G}(\boldsymbol{r};\boldsymbol{\hat\ell})]\cdot\boldsymbol{\hat r}\boldsymbol{\hat r}}=-(a/r^3)\boldsymbol{\mu}^\mathrm{tt}(\boldsymbol{\hat \ell})$. Combining with Eq. \eqref{eq:syl2} and insertion in Eq. \eqref{eq:syl} gives
\begin{equation}
\overline{\mathcal{L}_0\mathsfbi{G}(\boldsymbol{r};\boldsymbol{\hat\ell})\cdot\boldsymbol{\hat r}\boldsymbol{\hat r}}=\left(\frac{a}{2r}-\frac{a^3}{6r^3}\right)\boldsymbol{\mu}^\mathrm{tt}(\boldsymbol{\hat\ell}). \label{eq:Bogdan2}
\end{equation}
\end{appen}

\bibliographystyle{jfm}
\bibliography{jfm}



\end{document}